\documentclass[aps,prl,twocolumn,superscriptaddress]{revtex4-1} % 4.1 not installed

\usepackage{graphicx}
\usepackage{multirow}
\usepackage{amsmath}
\usepackage{color}

\newcommand{\ea}[1]{\begin{eqnarray}#1\end{eqnarray}}

\renewcommand{\eqref}[1]{(\ref{#1})}

\begin{document}

\title{Finite temperature correlations in the Lieb-Liniger 1D Bose gas}

\author{Mi{\l}osz Panfil}
\email[]{mpanfil@sissa.it}
%\thanks{}
\affiliation{Institute for Theoretical Physics, University of Amsterdam, Science Park 904\\
Postbus 94485, 1090 GL Amsterdam, The Netherlands}
\affiliation{International School for Advanced Studies (SISSA),\\
Via Bonomea 265, 34136, Trieste, Italy}

\author{Jean-S\'ebastien Caux}
\email[]{j.s.caux@uva.nl}
%\thanks{}
\affiliation{Institute for Theoretical Physics, University of Amsterdam, Science Park 904\\
Postbus 94485, 1090 GL Amsterdam, The Netherlands}

\date{\today}

\begin{abstract}

We address the problem of calculating finite-temperature response functions of an experimentally relevant low-dimensional strongly-correlated system: the integrable 1D Bose gas with repulsive $\delta$-function interaction (Lieb-Liniger model). Focusing on the dynamical density-density function, we present a Bethe Ansatz-based method allowing for its accurate evaluation in finite but large systems, over broad ranges of momenta, frequencies, temperatures and interaction parameters which are difficult to access using other methods. This allows us to quantify the reshaping of the zero temperature critical behavior by thermal fluctuations, in experimentally accessible regimes. 
\end{abstract}

%\keywords{}

\maketitle

Important examples of strongly correlated systems 
occur in reduced dimensionality \cite{GiamarchiBOOK}, in which the non-perturbative effects of interactions break any single-particle picture and can lead to quantum critical states. In particular, bosonic quantum gases confined to one-dimensional channels have recently been subjected to intense theoretical and experimental investigation \cite{2011_Cazalilla_RMP_83}. On the theoretical side, the physical responses of these systems, despite much progress, are still insufficiently understood to allow for high-quality experimental phenomenology. At low temperatures, 1D gases benefit from a universal Luttinger liquid description \cite{1981_Haldane_JPC_14} allowing to obtain the low-energy, long-distance asymptotics of observable correlations \cite{GiamarchiBOOK}. Alternately, methods based on integrability have allowed for the computation of ground state (zero temperature) dynamical correlations at arbitrary energy for continuum gases with contact interactions \cite{2006_Caux_PRA_74, *2007_Caux_JSTAT_P01008}. 

In experimental situations \cite{2006_Trebbia_PRL_97,2008_Amerongen_PRL_100,2012_Trotzky_NATPHYS_8,2011_Fabbri_PRA_83,2011_Jacqmin_PRL_106,2012_Armijo_PRL_108} thermal fluctuations cannot be discounted; since typical measurements (using {\it e.g.} Bragg spectroscopy \cite{2001_Brunello_PRA_64,2011_Fabbri_PRA_83}) require response functions away from the low-energy universal limit, the theoretical determination of correlations at finite temperature, energy and momentum scales is a crucial but difficult problem. In the context of the 1D Bose gas this has up to now only been partially addressed \cite{2005_Kheruntsyan_PRA_71,2006_Cherny_PRA_73,2009_Deuar_PRA_79,2009_Golovach_PRA_80,2010_Kormos_PRA_81,2011_Kozlowski_JSTAT_P03019,2012_Barthel_arXiv_12123570}.

In this letter we focus on the dynamical density-density response of the integrable Lieb-Liniger 1D Bose gas \cite{1963_Lieb_PR_130_1, *1963_Lieb_PR_130_2} at finite temperature. We present a Bethe Ansatz-based approach valid for interactions and temperatures covering physically interesting regimes. The nontrivial lineshapes obtained give quantitative predictions for eventual matching with experimental data. 

\paragraph{The model.-}
The Hamiltonian of the 1D Bose gas (Lieb-Liniger model \cite{1963_Lieb_PR_130_1, *1963_Lieb_PR_130_2}) is (setting $\hbar^2 = 2m = 1$)
\ea { \label{eq:H_LL}
H = -\sum_{i=1}^N \partial_{x_i}^2 + 2c\sum_{i>j}^N \delta(x_i - x_j) - \mu N,
} 
where $x_i$ denotes the position of the $i$-th atom and $\mu$ is the chemical potential. The coupling $c$ is related to the scattering length \cite{1998_Olshanii_PRL_81}. At finite temperatures, the gas is characterized by two  parameters: the interaction strength $\gamma = c/n$ and temperature $T$, where $n=N/L$ is the 1D density. Hereafter we set $n=1$ and $k_B = 1$. Hamiltonian \eqref{eq:H_LL} is exactly diagonalisable (in each sector of fixed particle number $N$) by Bethe Ansatz \cite{1963_Lieb_PR_130_1, *1963_Lieb_PR_130_2}. Imposing periodicity, eigenstates (labeled by quantum numbers $\{I_j\}_{j=1}^N$) 
are fully characterized by rapidities solving the Bethe equations \cite{1963_Lieb_PR_130_1}
\ea { \label{eq:Bethe}
\lambda_j + \sum_{k=1}^N\phi(\lambda_j-\lambda_k) = \frac{2\pi}{L}I_j,\;\;\;j=1,\dots, N.
} 
Here $\phi(\lambda) = 2\arctan(\lambda/c)$ is the 2-particle phase shift. The momentum and energy are 
\ea {
P_{\lambda} = \sum_{j=1}^N\lambda_j,\;\;\;\;E_{\lambda} = \sum_{j=1}^N \lambda_j^2.
}
The ground state is formed by a Fermi sea-like configuration of quantum numbers~\cite{1963_Lieb_PR_130_1}. Low-lying excitations can be classified in terms of particles and holes, these following their respective dispersion relations $\omega_{\pm}(k)$ \cite{1963_Lieb_PR_130_2}. At finite temperatures, the equilibrium state is (similarly to a free fermionic gas) a ``melted'' Fermi sea with smoothly-varying densities of particles and holes\cite{1969_Yang_JMP_10}. 
  
We are interested in dynamical properties in equilibrium at finite temperature. Although our method in principle applies to any few-point correlator, we focus on the experimentally-relevant density-density function 
\ea { \label{eq:DSF_eff}
  S_T(k,\omega) = \frac{2\pi}{L}\sum_{\lambda'} |\langle \lambda'| \hat{\rho}_k |\lambda_{\rho_T} \rangle|^2 \delta(\omega - E_{\lambda'} + E_{\rho_T}),
} 
where $|\rho_T\rangle$ is the thermal equlibrium state \cite{1969_Yang_JMP_10}, and the density operator is $\hat{\rho}(x) = \sum_{i=1}^N \delta(x - x_i)$. Its matrix element for any two eigenstates of the system and any value of the interaction parameter is known exactly \cite{1989_Slavnov_TMP_79, *1990_Slavnov_TMP_82} from Algebraic Bethe Ansatz \cite[and references therein]{KorepinBOOK}. 

The density-density correlation function at $T=0$ is characterized by a singular behavior along the dispersion lines $\omega_{\pm}(k)$ \cite{2008_Imambekov_PRL_100,2009_Imambekov_SCIENCE_323}. It vanishes below the lower dispersion $\omega_-(k)$ and has a power law singularity around $\omega_+(k)$. At small momentum and around umklapp excitations (with $K \approx 2mk_F$ and $\omega \approx 0$), this correlation is also singular with discontinuous support as usual for critical Luttinger liquids \cite{1981_Haldane_PRL_47}. We show later how these features are modified by thermal fluctuations.

Eq.~\eqref{eq:DSF_eff} is exact in the thermodynamic limit and finite-size corrections are of order of $1/L$\footnote{Note that experimentally, the situation is slightly altered by the presence of a trapping potential. This change can be understood within a local density approximation. The response of the 1D Bose gas was studied in \cite{2009_Golovach_PRA_80}; the trapping potential mostly affects low momenta, and remains small enough to justify studying the dynamics of homogeneous systems as a starting point.}. Their origin is two-fold: from the saddle-point approximation which was used to derive Eq.~\eqref{eq:DSF_eff} \cite{KorepinBOOK}, and from the evaluation of Eq. \eqref{eq:DSF_eff} in a finite system. We will quantify them later. 

Let us now discuss the evaluation of Eq. \eqref{eq:DSF_eff}.
We first obtain the distribution of rapidities in the thermal state following \cite{1969_Yang_JMP_10}. Choosing a fixed $N$, this was then approximated by a closest-matching state $|\{\lambda_T\}^N\rangle$. The representation \eqref{eq:DSF_eff} could then be scanned through the Hilbert space of relevant excitations (for this, the ABACUS algorithm \cite{2006_Caux_PRA_74, *2007_Caux_JSTAT_P01008} was extended to arbitrary excited states). Convergence was measured by the f-sum rule~\cite{LL_StatPhys2_BOOK}
\ea { \label{eq:fsum_rule}
  \int_{-\infty}^{\infty} \frac{d\omega}{2\pi}\omega S_T(k, \omega) = n k^2.
}
To verify that the computed correlation was indeed thermal, we used the f-sum rule combined with detailed balance ($S(k, \omega) = e^{-\beta\omega} S(k, -\omega)$)~\cite{LL_StatPhys2_BOOK}, yielding
\ea { \label{eq:fsum_rule+DB}
  \int_{0}^{\infty} \frac{d\omega}{2\pi}\omega S_T(k, \omega) \left(1 -e^{-\beta\omega}\right) = n k^2.
}
Repeating calculations for different system sizes then explicitly showed convergence to the thermodynamic limit.

\begin{figure}[ht]
  \includegraphics[scale = 0.4]{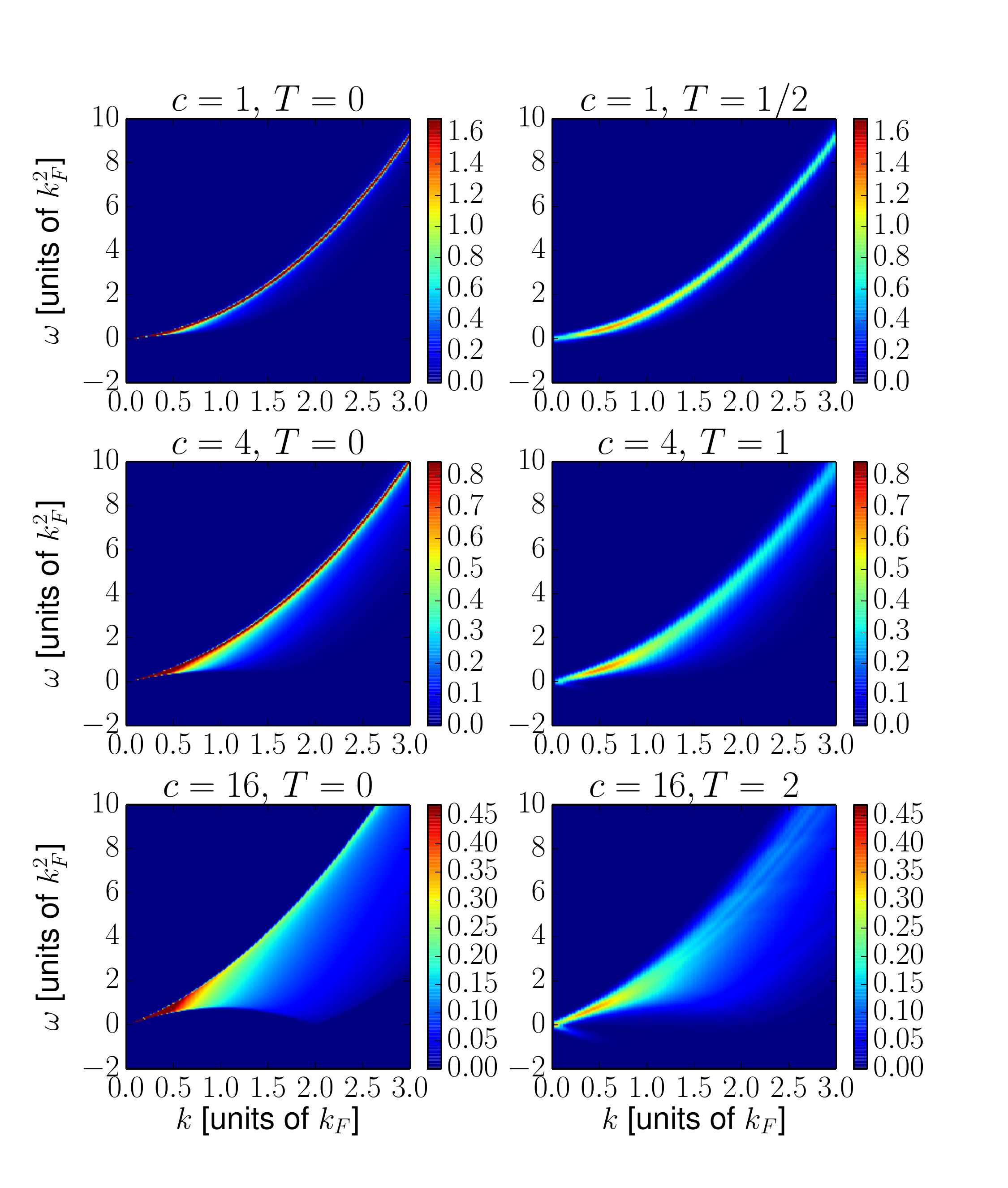}
  \caption{\label{fig:DSF}(\textit{color online}) The full dynamic correlator plotted for the intermediate values of interaction $c=1,4,16$ and for zero and high temperature. As the temperature increases the correlation becomes smeared but stays approximately within the same region in the $k-\omega$ plane. The exception being the small correlation region at low momentum and negative energy visible for $c=16$.}
\end{figure}

\begin{figure}[ht]
  \includegraphics[scale = 0.4]{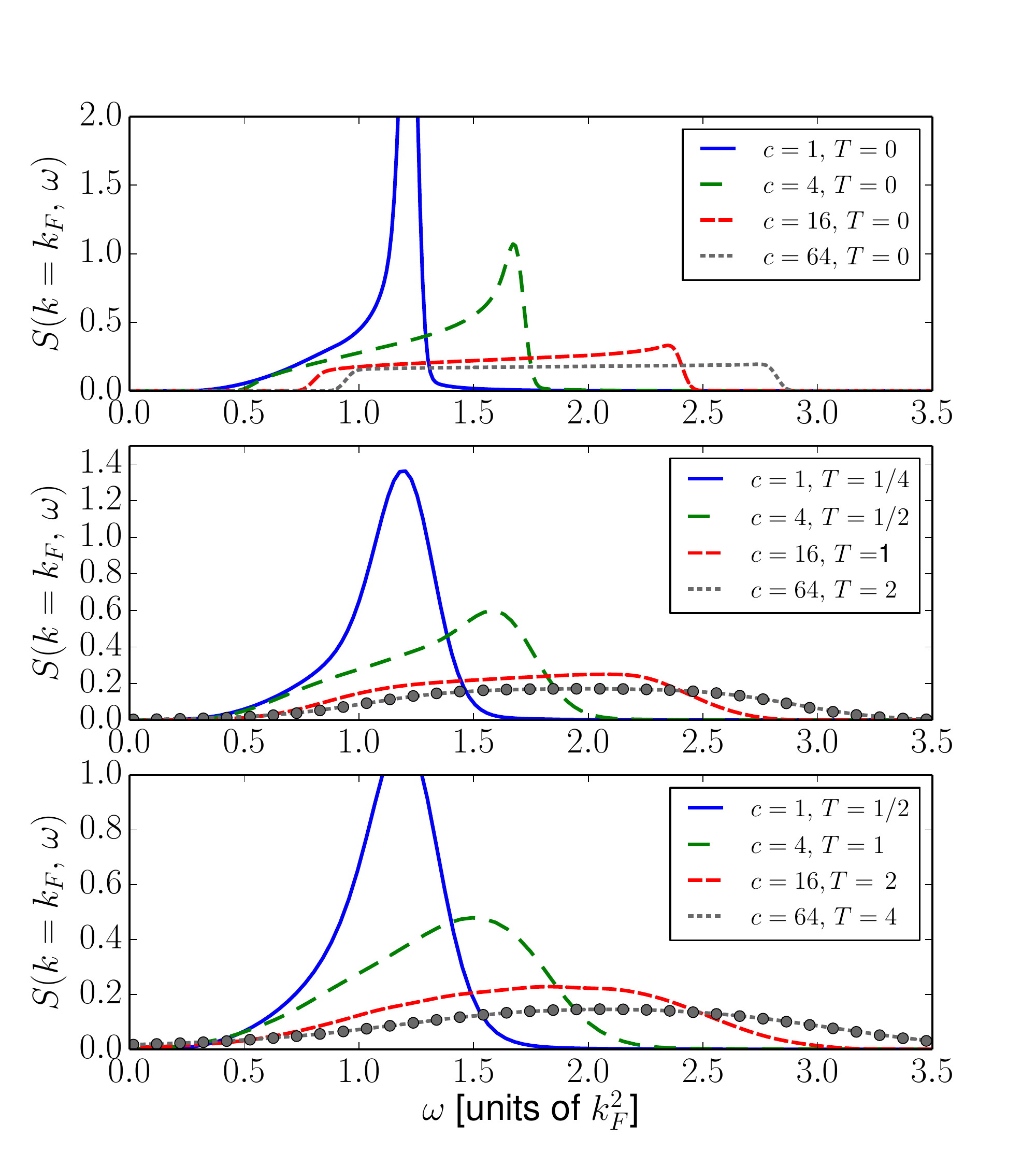}
  \caption{\label{fig:fixK_kF}(\textit{color online}) Fixed momentum cuts through the correlation function for $c=1, 4, 16, 64$ and at increasing values of temperatures from the top to the bottom. A finite temperature drastically modifies the lineshape of the correlation. The upper threshold singularities are washed out and the correlation becomes almost symmetric around its maximum. Results agree with the perturbative expansion in $1/c$ (dots) of \cite{2006_Cherny_PRA_73}. The errorbars, based on the f-sum rule (see Tab. \ref{tab:sum_rule}), are below the plot resolution (in the worst case they are around $1.5\%$; we include also the finite-size smoothening effects in this estimate)}
\end{figure}

\begin{figure}[ht]
  \includegraphics[scale=0.4]{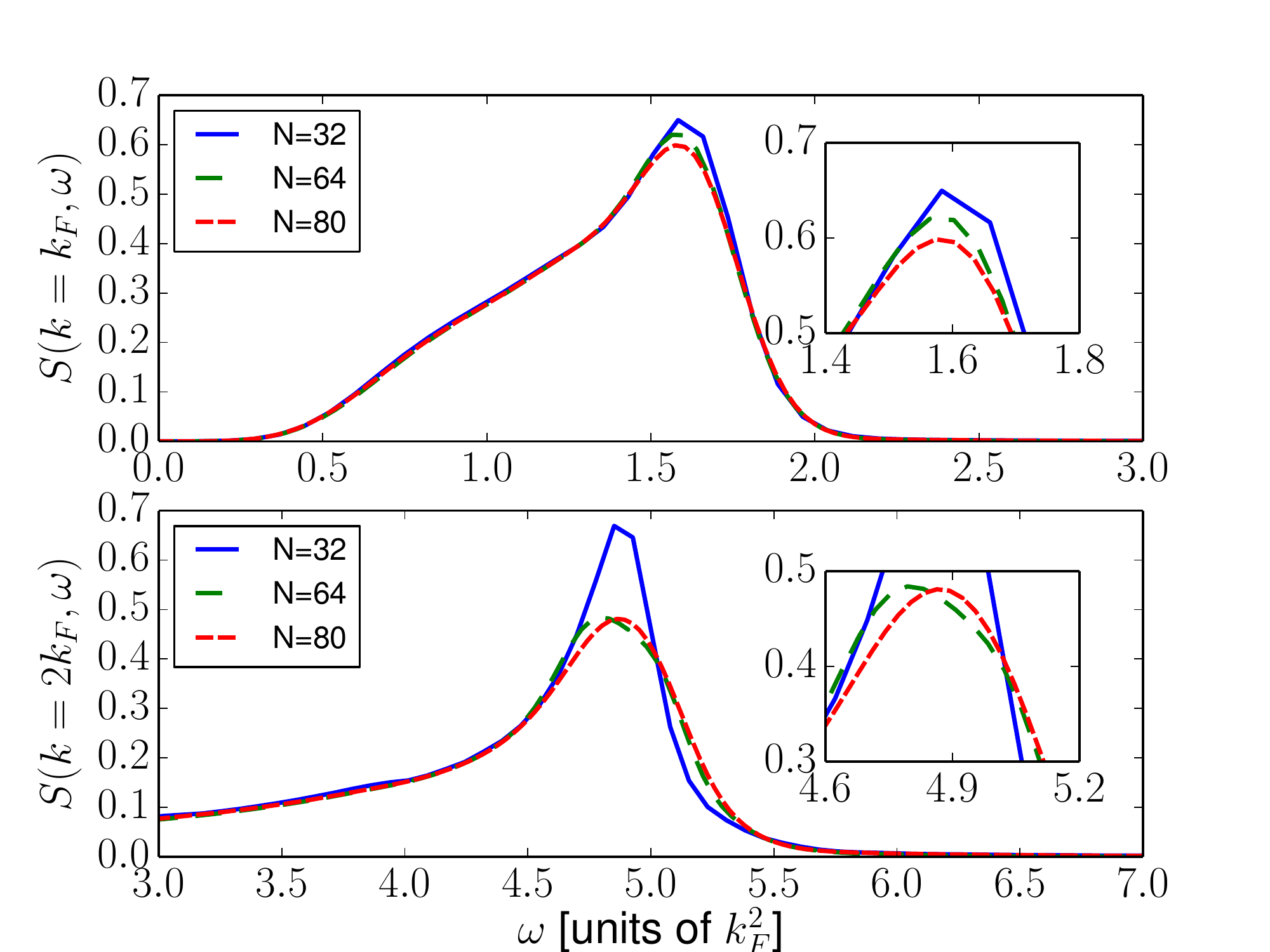}
  \caption{\label{fig:FS}(\textit{color online}) Fixed momentum cuts for $c=4$ and $T=0.5$ for 3 different system sizes explicitly showing convergence towards the thermodynamic limit. Insets contain zooms of the peak regions. Finite size effects are largest around $k=2k_F$ where the discrete nature of the thermal state is noticeable.}
\end{figure}

\paragraph{Results: momentum space.-}
The full $k$ and $\omega$ dependent density-density correlation function for various temperatures and interaction strengths is plotted in Fig. \ref{fig:DSF}. Representative f-sum rule saturations are presented in Tab. \ref{tab:sum_rule}. The $\omega$ dependence of the correlation is shown in Fig. \ref{fig:fixK_kF} where fixed momentum cuts (at $k=k_F$) are plotted. Fig.~\ref{fig:FS} illustrates finite-size effects.

\begin{table}
  \begin{tabular}{l l||c|c}
    & & $k=k_F$ & $k=2k_F$ \\\hline \hline
    \multirow{2}{*}{$c=1$} & $T = 1/4 \;(N=100)\;\;$ & $0.991$ & $0.975$ \\
    & $T = 1/2\;(N=64)$ & $0.991$ &  $0.979$ \\\hline
    \multirow{2}{*}{$c=4$} & $T=1/2\;(N=80)$ & $0.992$ & $0.982$ \\
    & $T=1\;(N=50)$ & $0.987$  &  $0.982$ \\\hline
    \multirow{2}{*}{$c=16\;$} & $T=1\;(N=100)$ & $0.990$ & $0.981$ \\
    & $T=2\;(N=64)$ & $0.997$ & $0.989$ \\\hline
  \end{tabular}
  \caption{\label{tab:sum_rule}The levels of saturation of the f-sum rule combined with the detailed balance relation (Eq. \eqref{eq:fsum_rule+DB}) for the intermediate interaction strengths and two values of momentum.}
\end{table}

The interacting 1D Bose gas has a single phase; however at finite interactions and temperatures we can still distinguish different regimes \cite{2003_Kheruntsyan_PRL_91,2008_Sykes_PRL_100,2009_Deuar_PRA_79}. For example, when interactions are strong and dominate over the temperature, one is in a fermionization regime. For weak interactions the gas resembles a quasi-condensate and can be analyzed using Bogolyubov theory with $T \ll \sqrt{c} \ll 1$. When the temperature dominates, the system resembles an ideal gas whose correlations can be obtained from a high-$T$ expansion \cite{2009_Deuar_PRA_79}. Here we set our attention on the experimentally relevant, but difficult-to-describe intermediate regime with $T/\sqrt{c} = 1/4, 1/2$ ($T=0$ curves being shown for reference).

The fixed momentum curves show the importance of thermal fluctuations in shaping the correlations. We begin by ascertaining the effect of temperature on the mean $\bar{\omega} = \int_0^{\infty} d\omega \omega S_T(k,\omega)$ and variance $\sigma^2 = \int_0^{\infty} d\omega \left(\omega - \bar{\omega}\right)^2 S_T(k, \omega)$. As seen from from Eq.~\eqref{eq:fsum_rule+DB}, the mean should increase with temperatures; similarly, the correlation should broaden. However, as can be seen from Tab.~\ref{tab:SA}, both effects are small. Even at finite temperatures, in the range we studied, the spread of the fixed momentum cuts is mainly due to interactions.

However, thermal fluctuations have an important and much more subtle effect in smoothening the singularities of the $T=0$ correlator. Two effects occur. First, a rounding off of the $T=0$ threshold singularities along the particle and hole modes $\omega_{\pm}(k)$. Second, a broadening of the correlation at very small momentum (also around umklapp excitations). The $T=0$ response in this limit is singular with vanishing width, and any thermal fluctuations destroy this feature. We come back to this when discussing the real space correlation function.

\begin{table}
  \begin{tabular}{l l||c|c}
    & & $\bar{\omega}(k_F)/\bar{\omega}_{T=0}(k_F)$ & $\sigma^2(k_F)/\sigma_{T=0}^2(k_F)$ \\\hline \hline
    \multirow{2}{*}{$c=1/4$} & $T = 1/8\;\;$ & $0.989$ & $1.023$ \\
    & $T = 1/4$ & $0.993$ &  $1.055$ \\\hline
    \multirow{2}{*}{$c=16$} & $T=1$ & $0.990$ & $1.033$ \\
    & $T=2$ & $0.995$  &  $1.115$ \\\hline
    \multirow{2}{*}{$c=256\;$} & $T=4$ & $1.006$ & $1.240$ \\
    & $T=8$ & $1.051$ & $1.595$ \\\hline
  \end{tabular}
  \caption{\label{tab:SA}Effects of temperature on the mean and variance of the positive energy part of the correlator (see main text for the definitions). We note that both mean and variance vary slightly while changing the temperature with the strongly interacting case ($c=256$) being an exception. The very slight decrease of the mean at lower temperatures is within the precision given by the f-sum rule (see Tab. \ref{tab:sum_rule}) and does not carry any physical meaning.}
\end{table}

\begin{figure}[ht]
  \includegraphics[scale = 0.4]{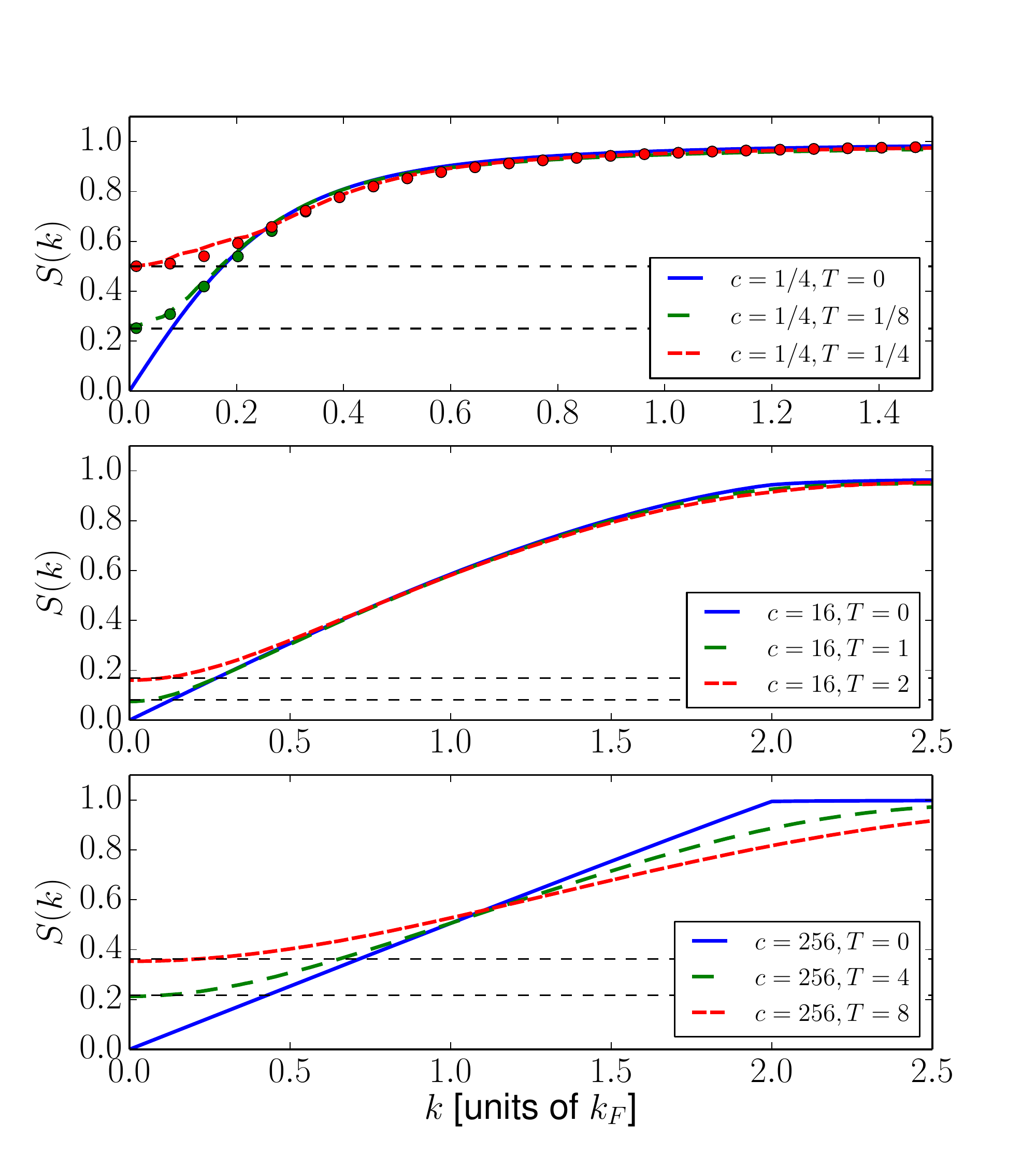}
  \caption{\label{fig:SSF}(\textit{color online}) Static structure factor for 3 representative values of the interaction strength ($c=1/4$, $c=16$, $c=256$). In the weakly interacting regime results agree with the Bogolyubov approximation (dots) \cite{BEC_BOOK}. The $k\rightarrow 0$ limit agrees with the hydrodynamic predictions Eq. \ref{eq:SSF_k0} (black dashed lines).}
\end{figure}

Integration over $\omega$ yields the static correlator (Fig. \ref{fig:SSF})
\ea { \label{eq:SSF}
  S(k) = \int_{-\omega}^{\omega} \frac{d\omega}{2\pi} S(k, \omega).
}
In the small momentum limit the dispersion relation of excitations becomes linear (\cite{LL_StatPhys2_BOOK, 1981_Haldane_PRL_47}, Fig. \ref{fig:DSF}) with the sound velocity given by the isothermal compressibility
%\ea {
$  v_s = \sqrt{2n\left(\frac{\partial \mu}{\partial n}\right)_T},$
%}
which can be calculated from the Thermodynamic Bethe Ansatz \cite{1969_Yang_JMP_10}. The f-sum rule combined with detailed balance then captures the correlation function, which becomes in this limit \cite{LL_StatPhys2_BOOK}
\ea { \label{eq:SSF_k0}
  S(0) = \left\{ 
  \begin{array}{l l}
    \frac{|k|}{v_s} + \mathcal{O}(k^2) & \quad T = 0,\\[1.2ex]
    \frac{2T}{v_s^2} + \mathcal{O}(k^2) & \quad T > 0.
  \end{array} \right.
}
The static correlator plotted in Fig. \ref{fig:SSF} agrees with this low momentum prediction and moreover, for weak interactions, confirms the validity of Bogolyubov theory.

\begin{figure}[ht]
  \includegraphics[scale = 0.4]{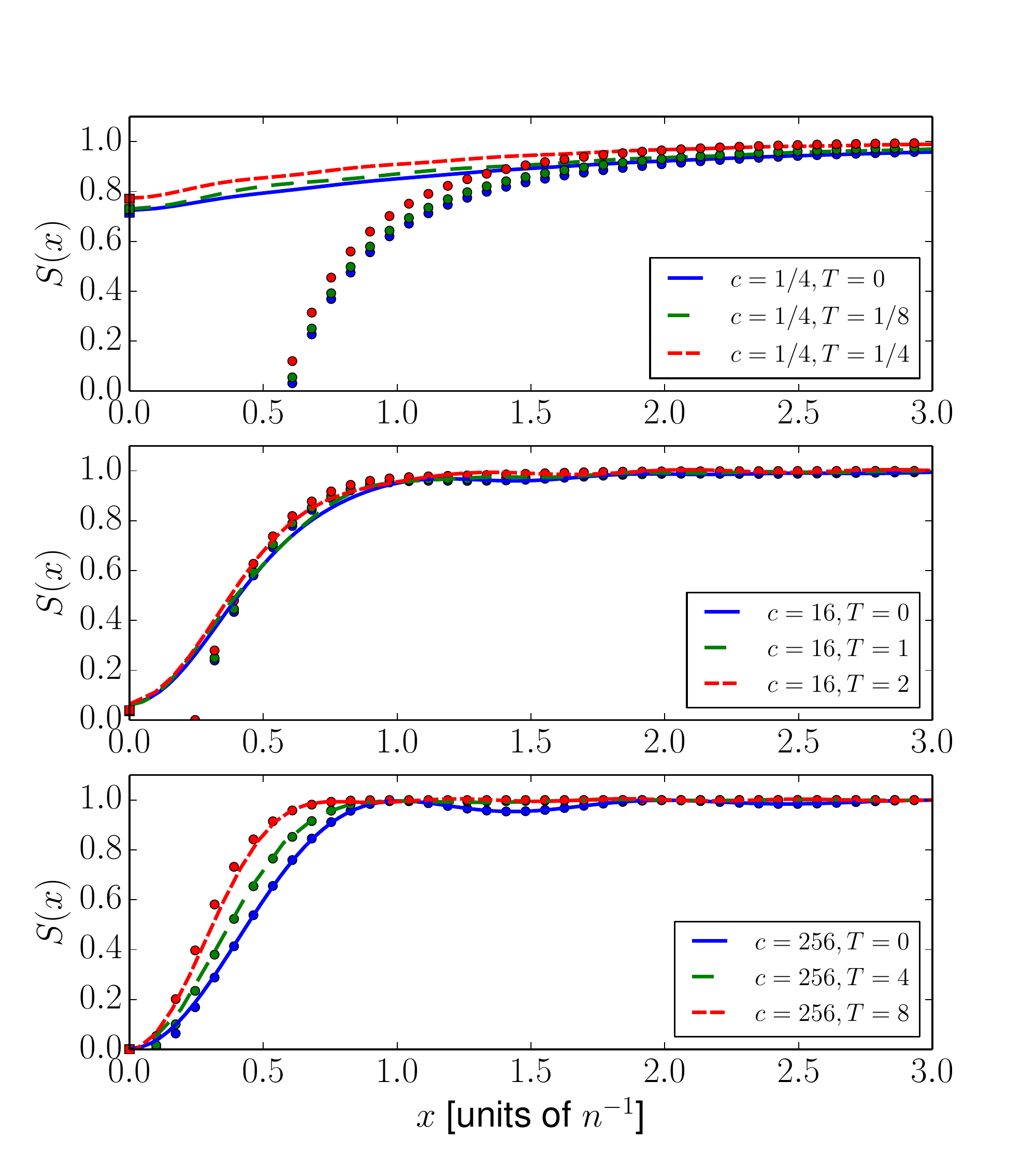}
  \caption{\label{fig:SFT}(\textit{color online}) The density-density correlation function in real space (see Eq. \eqref{eq:SFT}). The points are the Luttinger liquid predictions supplied with the zero-temperature prefactor $A$ (see Eq. \eqref{eq:SFT_LL} and discussion below it). The region of validity of the hydrodynamic predictions vary with the interaction strength and is the largest for $c\rightarrow\infty$. The value of the correlator at $x=0$ (squares) is calculated via the Hellmann-Feynmann theorem \cite{2003_Kheruntsyan_PRL_91, 2010_Kormos_PRA_81} and agrees with our predictions.}
\end{figure}

\paragraph{Results: real space.-}
The Fourier transform of the static correlator yields the pair correlation (Fig. \ref{fig:SFT})
\ea { \label{eq:SFT}
  S(x) = \frac{1}{L} \sum_k e^{-ikx} S(k).
}
Luttinger liquid theory \cite{1981_Haldane_JPC_14,1981_Haldane_PRL_47,GiamarchiBOOK} predicts exponential decay of this function at finite temperatures ($x \gg n^{-1}$)
\ea { \label{eq:SFT_LL}
  S_{LL}(x) &=& 1 - \frac{K}{2\pi^2} \left( \frac{\pi T/v_s}{\sinh(\pi Tx/v_s)}\right)^2 \nonumber\\
  &+& A \cos(2\pi x) \left( \frac{\pi T/v_s}{\sinh(\pi T x/v_s)}\right)^{2K} + \dots,
}
where $\dots$ represent terms that decay faster with distance, and where the prefactor $A$ is a non-universal number. The Luttinger parameter $K$ depends on the interaction strength $c$ and can be computed as in \cite{1981_Haldane_PRL_47}. For the 1D Bose gas at $T=0$, $A$ can be explicitly computed from the scaling limit of a single, specific matrix element of the density operator \cite{2011_Shashi_PRB_84, *2012_Shashi_PRB_85}. At finite temperature the relationship is more intricate; at low temperature the prefactor is however expected to be temperature independent \cite{2011_Kozlowski_JSTAT_P03019}. For the temperatures considered here (which go beyond the low-temperature limit) we find that the $T=0$ prefactor indeed gives predictions consistent with our results (see Fig. \ref{fig:SFT}). The correlation weight in the vicinity of the umklapp excitation is still the same (thus the same prefactor $A$) but is smeared over a finite region in energy and momentum. At $T=0$ this region shrinks to zero yielding a power-law decay instead of an exponential.

Throughout the manuscript we considered a homogeneous gas with a constant density of particles. In an experimental situation, where the presence of an external trapping potential leads to a spatially varying distribution of particles, the correlation function can be well approximated by fixing the density to an average density of particles in the trap.

\paragraph{Conclusions.-} 
In this paper we presented results for the finite temperature correlation function of the 1D Bose gas obtained through a combination of Bethe Ansatz and numerical evaluations of states and matrix elements. The results cover the experimentally-relevant regime of intermediate physical parameters (temperature, interaction, energy and momentum) which is difficult to access through other, analytical or numerical methods. We showed that for intermediate temperatures the correlation function carries remnants of $T=0$ characteristics such as signs of threshold singularities and exponential decay closely resembling the power-law decay. The exact lineshape of the correlation is however significantly and observably modified. The exact quantitative nature of our results should facilitate fitting with experimental predictions, perhaps paralleling what can be done for example in the context of ground state correlations in spin chains \cite{2013_Mourigal_NATPHYS_9}. In fact, besides extensions to other correlators, the method presented here is generalizable to other models solved by Bethe Ansatz, {\it e.g.} the XXZ spin chain. We will address this problem in future work.

\begin{acknowledgments}  
We thank C. Fort, N. Fabbri, L. Fallani, D. Cl\'ement, F.H.L. Essler and R. Konik for stimulating and fruitful discussions. 
We gratefully acknowledge support from the Foundation for Fundamental Research on Matter (FOM) and from the Netherlands Organisation for Scientific Research (NWO).
\end{acknowledgments}

\bibliography{/Users/milosz/WORK/LATEX/BIBTEX_LIBRARY_JSCaux_PAPERS,/Users/milosz/WORK/LATEX/BIBTEX_LIBRARY_JSCaux_BOOKS,/Users/milosz/WORK/LATEX/BIBTEX_LIBRARY_JSCaux_OWNPAPERS,/Users/milosz/WORK/LATEX/ExtraPAPERS}

%\bibliography{/Users/jscaux/WORK/BIBTEX_LIBRARY/BIBTEX_LIBRARY_JSCaux_PAPERS,/Users/jscaux/WORK/BIBTEX_LIBRARY/BIBTEX_LIBRARY_JSCaux_BOOKS,/Users/jscaux/WORK/BIBTEX_LIBRARY/BIBTEX_LIBRARY_JSCaux_OTHERS,/Users/jscaux/WORK/BIBTEX_LIBRARY/BIBTEX_LIBRARY_JSCaux_OWNPAPERS}

\end{document}